\documentclass[fleqn,twoside]{article}
\usepackage[headings]{espcrc2}
\readRCS $Id: espcrc2.tex,v 1.2 2004/02/24 11:22:11 spepping Exp $
\usepackage{graphicx}
\usepackage[figuresright]{rotating}

\def\beq{\begin{eqnarray}}
\def\eeq{\end{eqnarray}}
\def\bea{\begin{eqnarray}}
\def\eea{\end{eqnarray}}\def\be{\begin{equation}}
\def\ee{\end{equation}}\def\nn{\nonumber}
\newcommand{\spur}[1]{\not\! #1 \,}

\newcommand{\AmS}{{\protect\the\textfont2
  A\kern-.1667em\lower.5ex\hbox{M}\kern-.125emS}}

\title{Investigating quantum numbers of  new $c{\bar s}$ states }

\author{F. De Fazio\address{Istituto Nazionale di Fisica
Nucleare, Sezione di Bari, Italy}}

\begin{document}

\begin{abstract}
  \vspace{1pc}
  I discuss possible identification of the recently discovered
  mesons $D_{sJ}(2860)$ and $D_{sJ}(2700)$.

  This paper is dedicated to the memory of my professor, colleague and friend
  Beppe Nardulli.
\end{abstract}

\maketitle

\section{PREMISE}

The analysis of hadron properties may be simplified  exploiting
the symmetries that Quantum Chromodynamics (QCD) exhibits in
specific limits. An example is chiral $SU(N_f)_L \times SU(N_f)_R$
symmetry holding in the limit of $N_f$ massless quarks. This
symmetry is spontaneously broken to $SU(N_f)_V$ and light
pseudoscalar mesons are identified as Goldstone bosons acquiring
mass when explicit symmetry breaking mass terms are considered. An
effective theory (chiral perturbation theory) can be built as an
expansion in the light quark masses and momenta
\cite{Gasser:1984gg}.

Moreover, in the infinite heavy quark mass limit $m_Q \to \infty$,
the QCD Lagrangian is invariant under heavy quark spin and flavour
rotations. The corresponding effective theory is known as Heavy
Quark Effective Theory (HQET) \cite{HQET}.

Interactions of heavy mesons with light ones can be described by
an effective Lagrangian displaying both
 heavy quark symmetries, both  chiral symmetry.
 The approach was first formulated in the
 case of light pseudoscalars \cite{hqet_chir}, and later on it was extended to
  light vector mesons \cite{casalbuoni}.

 I would like to dedicate this paper to  Beppe
 Nardulli, who died on June 26th 2008. Beppe was full professor
  at Bari University teaching theoretical Physics. He
 gave an important contribution to the formulation of the effective Lagrangian
  describing heavy meson interactions with light mesons,
 as is summarized in a well known review paper
 \cite{Casalbuoni:1996pg}. More recently, he applied similar methods  to build
 an effective  theory valid in the high density regime of QCD
 \cite{Nardulli:2002ma}. He obtained important results  in
 this field as well \cite{Casalbuoni:2003wh}. It is impossible to summarize here all his
 contributions to Physics, which comprehend not only particle Physics,
  but also statistical mechanics and neural network Physics.
This
 paper is
 just my personal tribute to him. In the following I
discuss how the use  of effective Lagrangians
 for heavy mesons interacting with light pseudoscalars
 can shed light on new issues in charm spectroscopy.

\section{NEW $c \bar s$ MESONS }

Since 2003 there have been many new discoveries of open and hidden
charm hadrons \cite{reviews}. Here we focus on $c \bar s$  mesons.

Before the   B-factory era  the known $c \bar s$  spectrum
consisted of
   the pseudoscalar  $D_s(1968)$ and  vector $D_s^*(2112)$ mesons,
corresponding to   $s$-wave states of the  quark model, and the
axial-vector $D_{s1}(2536)$ and tensor $D_{s2}(2573)$ mesons,
$p$-wave states. In 2003,  two narrow  resonances: $D_{sJ}(2317)$
with $J^P=0^+$
 and $D_{sJ}^*(2460)$ with $J^P=1^+$ were discovered by BaBar \cite{Aubert:2003fg}
 and CLEO \cite{Besson:2003cp} Collaborations.
 Their
identification as  proper $c \bar s$ states was   debated
\cite{Colangelo:2004vu};  however, they have the quantum numbers
of the  states needed to complete the $p$-wave multiplet, and
their radiative decays occur accordingly, so that their
interpretation  as ordinary $c{\bar s}$ states is natural
\cite{fdf1}.

 In 2006,  BaBar
observed  another $c{\bar s}$ meson, $D_{sJ}(2860)$, decaying to
$D^0 K^+$ and $D^+ K_S$, with mass $M=2856.6 \pm 1.5 \pm 5.0$ MeV
and width $\Gamma=47 \pm 7 \pm 10$ MeV \cite{Aubert:2006mh}.
 Shortly after,
  analysing the $M^2(D^0 K^+)$  distribution in
 $B^+ \to {\bar D}^0 D^0 K^+$  Belle Collaboration
\cite{Brodzicka:2007aa}  established the presence of a $J^P=1^-$
resonance, $D_{sJ}(2700)$, with  $M = 2708\pm 9 ^{+11}_{-10}$ MeV
and $\Gamma= 108 \pm23 ^{+36}_{-31}$
 MeV.

Here we report a study aimed at properly identifying
$D_{sJ}(2860)$ and $D_{sJ}(2700)$. We analyze  their strong decays
 comparing the predictions that follow from different
assignments.

\section{DECAYS OF $D_{sJ}(2860)$ AND $D_{sJ}(2700)$ TO LIGHT PSEUDOSCALARS}

The study of mesons with a single heavy quark $Q$ is simplified in
the heavy quark limit $m_Q \to \infty$ when the spin $s_Q$ of the
heavy quark and  the angular momentum
 $s_\ell$ of the  light degrees of freedom: $s_\ell=s_{\bar q}+ \ell$ ($s_{\bar q}$
being the light antiquark spin and $\ell$ the orbital angular
momentum of the light degrees of freedom relative to $Q$) are
decoupled. Hence   spin-parity  $s_\ell^P$ of the light degrees of
freedom is conserved in strong interactions \cite{HQET} and mesons
can be classified as doublets of $s_\ell^P$. Two states with
$J^P=(0^-,1^-)$, denoted as $(P,P^*)$, correspond to $\ell=0$. The
four states corresponding to $\ell=1$ can be collected in two
doublets, $(P^*_{0},P_{1}^\prime)$ with  $ s_\ell^P={1 \over 2}^+$
and $J^P=(0^+,1^+)$, $(P_{1},P_{2})$ with $ s_\ell^P={3 \over
2}^+$ and $J^P=(1^+,2^+)$. For $\ell=2$ the doublets have
$s_\ell^P={3 \over 2}^-$, consisting of states with
$J^P=(1^-,2^-)$, or  $ s_\ell^P={5 \over 2}^-$  with
$J^P=(2^-,3^-)$ states. And so on.

 $D_s(1968)$, $D_s^*(2112)$ can be identified
with the members of the lowest lying $s_\ell^P={1 \over 2}^-$
doublet.
 $D_{s1}(2536)$,
$D_{s2}(2573)$, together with $D_{sJ}(2317)$,
  $D_{sJ}^*(2460)$,  fill the four $p$-wave levels: in particular, $D_{s2}(2573)$
 corresponds to
 $ s_\ell^P={3 \over 2}^+$,  $J^P=2^+$ state, while $D_{sJ}(2317)$
 to  $ s_\ell^P={1 \over 2}^+$,  $J^P=0^+$.
 The $J^P=1^+$ mesons $D_{s1}(2536)$ and $D_{sJ}^*(2460)$ could be  a mixing
   of   $ s_\ell^P={3 \over 2}^+$ and  $ s_\ell^P={1\over 2}^+$ states,  allowed
 at $O(1/m_Q)$; however, for non-strange charm mesons such a mixing was
  found to be small \cite{Abe:2003zm,Colangelo:2005gb}, so that we can identify $D_{s1}(2536)$ and $D_{sJ}^*(2460)$
 with the $J^P=1^+$  $ s_\ell^P={3 \over 2}^+$ and  $ s_\ell^P={1\over 2}^+$ states, respectively.

In the heavy quark limit,  the doublets are represented by
effective fields: $H_a$ for $s_\ell^P={1\over2}^-$ ($a=u,d,s$ is a
light flavour index), $S_a$ and $T_a$ for $s_\ell^P={1\over2}^+$
and $s_\ell^P={3\over2}^+$, respectively; $X_a$ and $X^\prime_a$
for
  $s_\ell^P={3\over2}^-$ and $s_\ell^P={5\over2}^-$, respectively:
 \bea
&& \hskip -0.6cm H_a  = \frac{1+{\rlap{v}/}}{2}[P_{a\mu}^*\gamma^\mu-P_a\gamma_5]  \label{neg} \nn  \\
&&  \hskip -0.6cm S_a = \frac{1+{\rlap{v}/}}{2} \left[P_{1a}^{\prime \mu}\gamma_\mu\gamma_5-P_{0a}^*\right]   \nn \\
&&  \hskip -0.6cm T_a^\mu=\frac{1+{\rlap{v}/}}{2} \Bigg\{
P^{\mu\nu}_{2a} \gamma_\nu \label{pos2}  \\
&&  \hskip  0.6cm -P_{1a\nu} \sqrt{3 \over 2} \gamma_5
\left[g^{\mu \nu}-{1 \over 3} \gamma^\nu (\gamma^\mu-v^\mu)
\right]
\Bigg\} \nn  \\
&&  \hskip -0.6cm X_a^\mu =\frac{1+{\rlap{v}/}}{2} \Bigg\{
P^{*\mu\nu}_{2a} \gamma_5 \gamma_\nu \nn  \\
&&  \hskip  0.6cm -P^{*\prime}_{1a\nu} \sqrt{3 \over 2}  \left[
g^{\mu \nu}-{1 \over 3} \gamma^\nu (\gamma^\mu-v^\mu) \right]
\Bigg\}   \nn   \\
&&  \hskip -0.6cm X_a^{\prime \mu\nu} =\frac{1+{\rlap{v}/}}{2}
\Bigg\{ P^{\mu\nu\sigma}_{3a} \gamma_\sigma
 -P^{*'\alpha\beta}_{2a} \sqrt{5 \over 3}
\gamma_5 \Bigg[ g^\mu_\alpha g^\nu_\beta
\nn  \\
&&  \hskip  0.6cm-{1 \over 5} \gamma_\alpha g^\nu_\beta
(\gamma^\mu-v^\mu) -{1 \over 5} \gamma_\beta g^\mu_\alpha
(\gamma^\nu-v^\nu) \Bigg] \Bigg\} \nn \eea with the various
operators annihilating mesons of four-velocity $v$ (conserved in
strong interactions) and   containing a factor $\sqrt{m_P}$. Light
pseudoscalars are introduced using
 $\displaystyle \xi=e^{i {\cal M} \over
f_\pi}$,with: \bea  \small {\cal M}= \left(\begin{array}{ccc}
\sqrt{\frac{1}{2}}\pi^0+\sqrt{\frac{1}{6}}\eta & \pi^+ & K^+\nonumber\\
\pi^- & -\sqrt{\frac{1}{2}}\pi^0+\sqrt{\frac{1}{6}}\eta & K^0\\
K^- & {\bar K}^0 &-\sqrt{\frac{2}{3}}\eta
\end{array}\right)\nn
\eea ($f_{\pi}=132 \; $ MeV).  At the leading order in the heavy
quark mass and light meson momentum expansion the decays  $F \to H
M$ $(F=H,S,T,X,X^\prime$ and $M$ a light pseudoscalar meson) can
be described by the Lagrangian interaction  terms (invariant under
chiral and
 heavy-quark spin-flavour transformations)
\cite{hqet_chir,casalbuoni}:
\bea && \hskip -0.6cm {\cal L}_H = \,  g \, Tr [{\bar H}_a H_b
\gamma_\mu \gamma_5 {\cal
A}_{ba}^\mu ] \nn \\
&& \hskip -0.6cm {\cal L}_S =\,  h \, Tr [{\bar H}_a S_b
\gamma_\mu \gamma_5 {\cal
A}_{ba}^\mu ]\, + \, h.c. \,,  \label{lag-hprimo} \\
&& \hskip -0.6cm {\cal L}_T =  {h^\prime \over
\Lambda_\chi}Tr[{\bar H}_a T^\mu_b (i D_\mu {\spur {\cal
A}}+i{\spur D} { \cal A}_\mu)_{ba} \gamma_5
] + h.c.   \nn  \\
&& \hskip -0.6cm {\cal L}_X =  {k^\prime \over
\Lambda_\chi}Tr[{\bar H}_a X^\mu_b (i D_\mu {\spur {\cal
A}}+i{\spur D} { \cal A}_\mu)_{ba} \gamma_5
] + h.c.  \nn \\
&& \hskip -0.6cm {\cal L}_{X^\prime} =  {1 \over
{\Lambda_\chi}^2}Tr[{\bar H}_a X^{\prime \mu \nu}_b [k_1 \{D_\mu,
D_\nu\} {\cal A}_\lambda  \nn \\
&& \hskip 0.4cm + k_2 (D_\mu D_\nu { \cal A}_\lambda + D_\nu
D_\lambda { \cal A}_\mu)]_{ba}  \gamma^\lambda \gamma_5] + h.c.
\nn \eea
where  $D_{\mu ba}
=-\delta_{ba}\partial_\mu+\frac{1}{2}\left(\xi^\dagger\partial_\mu
\xi +\xi\partial_\mu \xi^\dagger\right)_{ba}$, ${\cal A}_{\mu
ba}=\frac{i}{2}\left(\xi^\dagger\partial_\mu \xi-\xi
\partial_\mu \xi^\dagger\right)_{ba}$. $\Lambda_\chi \simeq 1 \, $ GeV
is  the chiral symmetry-breaking scale. $g, h$, $h^\prime$,
$k^\prime$, $k_1$ and $k_2$ represent  effective coupling
constants.

The structure of the Lagrangian terms for radial excitations of
$H$, $S$ and $T$ does not change, but $g, h$, $h^\prime$ should be
substituted by $\tilde g, \tilde h$, $\tilde h^\prime$.

Let us start with $D_{sJ}(2860)$. A new $c \bar s$ meson decaying
to $DK$ can be either  the $J^P=1^-$ state of the  $ s_\ell^P={3
\over 2}^-$ doublet,  or the $J^P=3^-$ state of the  $ s_\ell^P={5
\over 2}^-$ one, in both cases with  lowest radial quantum number.
Otherwise $D_{sJ}(2860)$ could be a radial excitation of already
observed $c\bar s$ mesons: the first radial excitation of $D_s^*$
($J^P=1^-$   $ s_\ell^P={1 \over 2}^-$) or of $D_{sJ}^*(2317)$
($J^P=0^+$   $ s_\ell^P={1 \over 2}^+$) or  of $D_{s2}(2573)$
($J^P=2^+$ $ s_\ell^P={3 \over 2}^+$).

As for $D_{sJ}(2700)$, two possibilities  can   be considered, the
spin having been already fixed:
 {\it i)} $D_{sJ}(2700)$ belongs to the $ s_\ell^P={1
\over 2}^-$ doublet    and is the first radial excitation ($D_s^{*
\prime}$); {\it ii)} $D_{sJ}(2700)$ is the low lying state with $
s_\ell^P={3 \over 2}^-$ ($D_{s1}^{* }$).

In \cite{Colangelo:2006rq}  we investigated  the decay modes of
$D_{sJ}(2860)$ and $D_{sJ}(2700)$ according to the various
possible assignments with the aim of discriminating among them.
The results are collected in Table \ref{ratios} where we report
the ratios $R_1=\displaystyle {\Gamma( D_{sJ}  \to D^*K) \over
\Gamma( D_{sJ} \to DK) }$ and  $R_2=\displaystyle {\Gamma( D_{sJ}
\to D_s \eta) \over \Gamma( D_{sJ} \to DK) }$ (with
$D^{(*)}K=D^{(*)+} K_S +D^{(*)0} K^+$) obtained for various
quantum number assignments to $D_{sJ} (2860)$ and $D_{sJ}(2700)$
using
 eqs.(\ref{pos2}) and (\ref{lag-hprimo}). The ratios
 do not depend on the coupling constants,  but only on the quantum numbers.
%
\begin{table}[bt]
\caption{Predicted ratios
    $R_1$ and $R_2$ (see text for definitions)
 for the various assignment
 of quantum numbers to  $D_{sJ}(2860) $ and $D_{sJ}(2700)$.  }
    \label{ratios}
    \begin{center}
    \begin{tabular}{| c |  c | c |}
       \hline
 $D_{sJ}(2860) $  &$R_1$  &   $R_2$
\\ \hline
 $s_\ell^p={1\over 2}^-$, $J^P=1^-$,  $n=2$  &$1.23$& $0.27$ \\
$s_\ell^p={1\over 2}^+$, $J^P=0^+$, $n=2$   &$0$& $0.34$ \\
$s_\ell^p={3\over 2}^+$, $J^P=2^+$, $n=2$   &$0.63$& $0.19$\\
$s_\ell^p={3\over 2}^-$, $J^P=1^-$,   $n=1$   & $0.06$& $0.23$ \\
$s_\ell^p={5\over 2}^-$, $J^P=3^-$,   $n=1$   & $0.39$& $0.13$ \\
    \hline \hline $D_{sJ}(2700) $  &$R_1$  &   $R_2$
\\ \hline $s_\ell^p={1\over 2}^-$, $J^P=1^-$,  $n=2$ & $91 $ & $20 $   \\ \hline
  $s_\ell^p={3\over 2}^-$, $J^P=1^-$,   $n=1$ & $4.3 $ & $16.3 $  \\
  \hline
    \end{tabular}
  \end{center}
\vskip -1 cm\end{table}
We first discuss  the entries in Table \ref{ratios} concerning
$D_{sJ}(2860)$. In particular, non observation (at present)  of a
$D^*K$ signal in the $D_{sJ}(2860)$ range of mass implies that the
production of $D^* K$ is not favoured,  therefore  the assignments
$s_\ell^p={1\over 2}^-$, $J^P=1^-$, $n=2$,  and $s_\ell^p={3\over
2}^+$, $J^P=2^+$, $n=2$ can be excluded.

The case $s_\ell^p={3\over 2}^-$, $J^P=1^-$, $n=1$ can also be
excluded since,   using the relevant  term  in (\ref{lag-hprimo})
and $k^\prime \simeq h^\prime\simeq 0.45\pm0.05$ (as  the
$h^\prime$   was determined in \cite{Colangelo:2005gb}),
 would give  $\Gamma(D_{sJ}\to DK)\simeq 1.5$ GeV, a result
 incompatible with the measured width.

In the  assignment  $s_\ell^p={1\over 2}^+$, $J^P=0^+$, $n=2$ the
decay to $D^*K$ is forbidden.  However, if $D_{sJ}(2860)$ is a
scalar radial excitation, it should have  a spin partner with
 $J^P=1^+$ ($s_\ell^p={1\over 2}^+$, $n=2$)  decaying to $D^* K$ with a small width,
  a rather easy signal to detect.
 For $n=1$ both  $D^*_{sJ}(2317)$ and $D_{sJ}(2460)$ are produced in charm continuum
 at  $e^+ e^-$ factories.
To explain the absence of the $D^*K$ in charm continuum events  at
mass around $2860$ MeV, one should  invoke some mechanism
favouring the production of the $0^+$ $n=2$ state and inhibiting
the production of $1^+$ $n=2$ state,
 a mechanism which discriminates the first radial excitation from the low lying state $n=1$.
 Such a mechanism is  difficult to imagine.
 \footnote{The
interpretation of $D_{sJ}(2860)$ as the first radial excitation of
$D^*_{sJ}(2317)$  has been proposed in  \cite{vanBeveren:2006st}.}

The last possibility is:  $s_\ell^p={5\over 2}^-$, $J^P=3^-$,
$n=1$. In this case, the small $DK$ width is due to the huge
suppression related to the kaon momentum factor: $\displaystyle
\Gamma(D_{sJ}\to DK) \propto  q_K^7$. The spin partner would be
$D_{s2}^{*}$, the $s_\ell^P={5 \over 2}^-$, $J^P=2^-$ state,
 which can  decay to $D^* K$ and not to $DK$. It
would also  be narrow but  only  in the $m_Q\to \infty$ limit,
where the transition  $D_{s2}^{*}\to D^*K$ occurs in $f$-wave. As
an effect of $1/m_Q$ corrections this  decay can occur in
$p$-wave, so that   $D_{s2}^{*}$ could be  broader; therefore,  it
is not necessary to invoke a mechanism inhibiting the production
of this state with respect to $J^P=3^-$. If  $D_{sJ}(2860)$ has
$J^P=3^-$, it is not expected to  be produced
 in non leptonic $B$ decays such as
$B \to  D D_{sJ}(2860)$: the non leptonic amplitude in the
factorization approximation vanishes   since the vacuum matrix
element of the weak $V-A$ current with a spin three particle is
zero. Therefore,  the quantum number assignment can be confirmed
by studies of $D_{sJ}$ production in $B$ transitions. Actually, in
the Dalitz plot analysis of $B^+ \to \bar D^0 D^0 K^+$ Belle
Collaboration \cite{Brodzicka:2007aa} has reported  no signal of
$D_{sJ}(2860)$.

The conclusion  of our study  is that $D_{sJ}(2860)$ is likely  a
$J^P=3^-$ state,
 a  predicted high mass and relatively narrow $c \bar s$ state
 \cite{Colangelo:2000jq}.

We now consider $D_{sJ}(2700)$. As  Table \ref{ratios} shows,
$R_1$ is very different if $D_{sJ}(2700)$ is $D_s^{*\prime}$ or
$D_{s1}^*$:
  the  $D^*K$  mode is the main  signal to be investigated in order to
distinguish between the two  possible assignments. From the
computed widths, assuming that  $\Gamma(D_{sJ}(2700))$ is
saturated by modes with a heavy meson and a light pseudoscalar
meson in the final state, we can determine the couplings $\tilde
g$ and $k^\prime$ governing the decays in the two cases.
Identifying $D_{sJ}(2700)$ with $D_s^{*\prime}$ we obtain: $\tilde
g= 0.26 \pm 0.05$ while if
 $D_{sJ}(2700)$ is $D_{s1}^*$ we get $k^\prime=0.14 \pm
0.03$. These  values are similar to those obtained for analogous
couplings  appearing in the effective heavy quark chiral
Lagrangians  \cite{Colangelo:1995ph}.

The results for $\tilde g$ and $k^\prime$ can provide information
about the spin partner of $D_{sJ}(2700)$, i.e. the state belonging
to the same  $s_\ell^P$ doublet   from which $D_{sJ}(2700)$
differs only for  the total spin. The partner of $D_s^{*\prime}$
($s_\ell^P={1 \over 2}^-$) has $J^P=0^-$; it is denoted
$D_s^\prime$,   the first radial excitation of $D_s$, while the
partner of $D_{s1}^*$ ($s_\ell^P={3 \over 2}^-$) is the state
$D_{s2}^*$ with $J^P=2^-$. In both cases,  the decay modes to $
D^{*0} K^+$, $ D^{*+} K^0_{S(L)}$, $ D^{*}_s \eta$, are permitted.
In the heavy quark limit, these partners are degenerate.  Using
the obtained values for $\tilde g$ and $k^\prime$,  we get:
$\Gamma(D_s^\prime)= (70 \pm 30)$  MeV and $\Gamma(D_{s2}^*)= (12
\pm 5)$ MeV,  so that in the two assignments the spin partners
differ for their  total width.

\section{CONCLUSIONS}

Studying decay rates of $D_{sJ}(2860)$  to light pseudoscalar
mesons we conclude that  most likely $D_{sJ}(2860) $ has
$J^P=3^-$. The detection of the final state $D^*K$ would support
this interpretation. As for $D_{sJ}(2700)$,  the decay mode to
$D^*K$ has very different branching ratios in the two possible
assignments, so that measuring such a branching fraction could
help to identify $D_{sJ}(2700)$.

\end{document}